\documentclass{article}
\usepackage{PRIMEarxiv}

\usepackage[utf8]{inputenc} 
\usepackage[T1]{fontenc}    
\usepackage{hyperref}       
\usepackage{url}            
\usepackage{booktabs}       
\usepackage{amsfonts}       
\usepackage{nicefrac}       
\usepackage{microtype}      
\usepackage{lipsum}
\usepackage{graphicx}
\usepackage{wrapfig}
\usepackage{authblk}
\graphicspath{{media/}}     

\begin{document}

\title{FACTS: Facial Animation Creation using the Transfer of Styles}

\author[1, 2]{Jack R. Saunders}
\author[2]{Steven J. Caulkin}
\author[1]{Vinay P. Namboodiri}
\affil[1]{Department of Computer Science, University of Bath. \{jrs68, vpn22\}@bath.ac.uk}
\affil[2]{Epic Games R\&D. steve.caulkin@cubicmotion.com}

\maketitle

\begin{figure*}[!h]
    \centering
    \includegraphics[width=0.9\columnwidth]{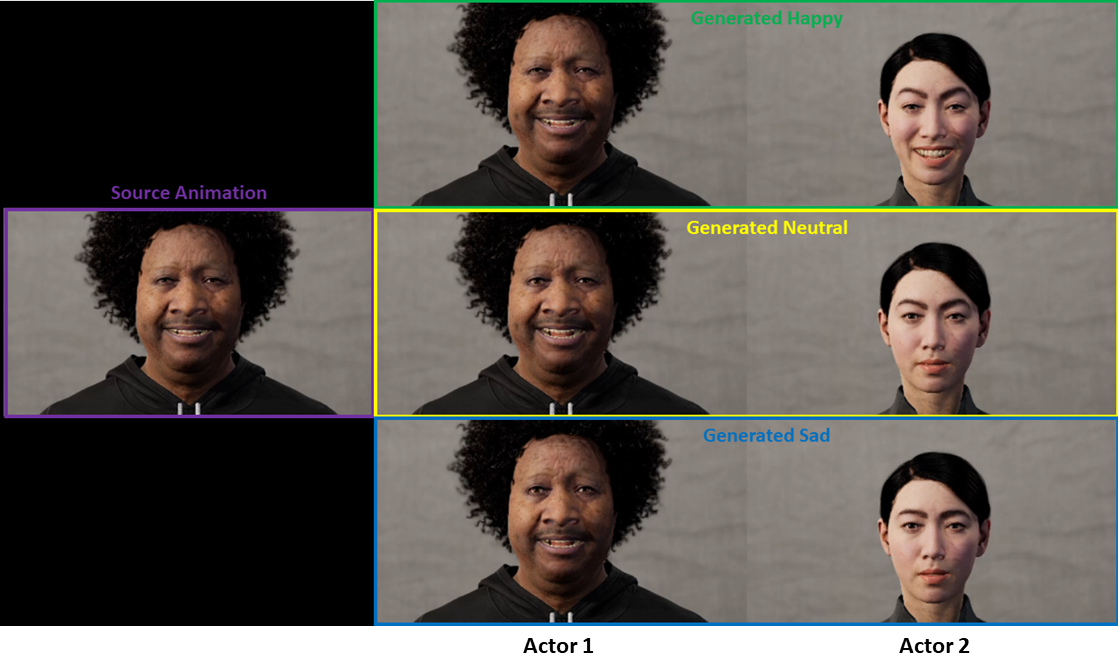}
    \caption{Our proposed method, FACTS is a model that takes an animation and a target style and produces a new animation in the desired style. This model is capable of altering both emotion and idiosyncratic style while preserving gestures and lip sync. This is achieved through a combination of loss functions, including a novel viseme-preserving loss.}
\end{figure*}

\begin{abstract}
    The ability to accurately capture and express emotions is a critical aspect of creating believable characters in video games and other forms of entertainment. Traditionally, this animation has been achieved with artistic effort or performance capture, both requiring costs in time and labor. More recently, audio-driven models have seen success, however, these often lack expressiveness in areas not correlated to the audio signal. In this paper, we present a novel approach to facial animation by taking existing animations and allowing for the modification of style characteristics. Specifically, we explore the use of a StarGAN to enable the conversion of 3D facial animations into different emotions and person-specific styles. We are able to maintain the lip-sync of the animations with this method thanks to the use of a novel viseme-preserving loss.
\end{abstract}

\section{Introduction}

Digital human characters are a crucial aspect of the entertainment industry, and there is an increasing demand for high-quality facial animations. However, traditional methods of facial animation, such as key-framing, are time-consuming and require significant artistic skill. As a result, new approaches are needed to generate large quantities of high-quality facial animations more efficiently. One popular alternative to traditional key-framing is the use of performance capture, which automatically infers an animation from recorded footage of an actor. However, performance capture is still a time-consuming process that requires a team of experts and trained actors, and the equipment can be expensive and complicated. While some attempts have been made to democratize this process by allowing for performance capture with consumer-grade devices, these methods still have limitations. Another approach to facial animation is the generation of animations from audio, which requires only speech as input and produces animations. While such methods are in their infancy, they often suffer from a lack of quality due to suppressing motion not directly correlated with audio.

We propose an alternative solution for generating animations that involves taking existing animations and altering certain style characteristics to create new animations. For example, if a developer is creating a new video game with hundreds of interactive characters, it would only be necessary to produce animations for a fraction of the characters, and by altering the style, unique animations could be produced for all of them.

In this work, we define two types of style: emotional and idiosyncratic. Emotional style refers to the changes in a character's facial expressions and movements that correspond to different emotional states, while idiosyncratic style refers to the unique mannerisms of a character. However, our proposed method should generalize to arbitrary styles, as long as there is sufficient diverse data available for that particular style.

We propose a many-to-many style transfer method using a modified StarGAN\footnote{We use the framework of StarGAN trained from scratch on our own data, we do not use a pre-trained model.} \cite{StarGAN2018}. We employ cycle-consistency \cite{CycleGAN2017, DiscoGAN2017} to learn style mappings without paired data, resulting in a more efficient and flexible approach to animation production. Additionally, we ensure the temporal consistency of our animations by treating rig controls as a time series and incorporating temporal information using a GRU \cite{GRU14} layer. We also introduce a novel viseme preservation loss, which significantly improves lip-sync to produce high-quality animations that are synchronized with audio. In summary, the contributions of this work are:

\begin{itemize}
    \item{A methodology for producing new animations by altering style characteristics.}
    \item{An adaptation of the StarGAN that works for animation data and can alter multiple styles with a single network.}
    \item{A novel viseme preserving loss that enables lip-sync without constraining the expressiveness of the mouth.}
\end{itemize}

\section{Related Work}

\subsection{Image Style Transfer}

Image style transfer is a widely studied problem \cite{pix2pix2017, CycleGAN2017, DiscoGAN2017, gatys2016image, StarGAN2018}, with early methods like Pix2Pix \cite{pix2pix2017} able to generate impressive results but requiring paired training data, which can be difficult to obtain for different styles of the same content. The cycle-consistency loss \cite{CycleGAN2017, DiscoGAN2017} was introduced to enable style transfer without paired data. This loss is calculated by applying a style transformation to an image, inverting the transformation, and comparing the reconstructed image to the original. Models trained with this loss learn to change only those image features that are correlated with the style while retaining other aspects of the original image.

Despite the success of the cycle-consistency loss, models like CycleGAN suffer from the limitation of only being able to perform one-to-one style mappings, meaning that a full mapping between $N$ different styles requires training $O(N^2)$ networks. StarGAN \cite{StarGAN2018} addresses this limitation by using a single network to perform many-to-many style mappings. In StarGAN, the generator takes an input image and a target style code to output a new image with the desired style. A discriminator is trained not only to distinguish real and fake images but also to classify the style. The generator is then encouraged to produce images classified as the target style by the discriminator. Interestingly, the authors found that StarGAN outperforms CycleGAN even for one-to-one style mappings, likely due to regularisation from multi-task learning. Given these advantages, we use the StarGAN framework for our animation style transfer method.

\subsection{Style Transfer in Facial Animation}

While style transfer in the image domain has seen significant progress, few works have adapted it to facial animation \cite{NSPVD, 3D-TalkEmo}. One notable paper in this area is Neural Style Preserving Visual Dubbing (NSPVD) \cite{NSPVD}, which shares similarities with our work. NSPVD performs style transfer over 3DMM parameters \cite{Blanz99, 3DMM2020} using a CycleGAN with an LSTM layer to capture temporal dynamics. A neural renderer is then used to preserve the target actor's idiosyncratic style. However, the CycleGAN backbone is only capable of one-to-one style mappings, which means that a new network must be trained for every pair of source and target styles. In contrast, our approach builds on the StarGAN \cite{StarGAN2018} framework and can handle many-to-many style mappings. Additionally, NSPVD uses constraints on the shape of the mouth to preserve lip-sync, which can cause issues in cases where the source and target styles have significantly different mouth shapes. Our proposed viseme-preserving loss, on the other hand, can maintain important mouth shapes without unnecessarily constraining expressions.

Another related work is the 3D-TalkEmo \cite{3D-TalkEmo} model, which uses a StarGAN to augment emotional data. However, this work is limited to static meshes and ignores temporal dynamics, leading to temporal inconsistencies. In contrast, our approach uses recurrent layers to capture the temporal dynamics of facial animation, preventing such inconsistencies.

\subsection{Audio-Driven Animation}

Audio-driven animation refers to a collection of methods that allow for the creation of facial animation from audio-only. While our proposed method is not an audio-driven one, it does share some similarities, in particular the emphasis on synchronising the lip animation to audio. Karras et. al. \cite{Karras17} directly regress vertex positions from audio using convolutional neural networks. They then learn additional latent vectors to represent all the features that are not directly correlated with audio, including emotion. These vectors lack semantic meaning however and so are `mined' to produce human-understandable labels. This mining process is highly manual. Furthermore, this method cannot model motion that is neither correlated to audio or emotion, such as eye blinks and is trained for a single person only. VOCA \cite{VOCA2019} improves generalisation over different identities by explicitly including a one-hot vector representation of identity. Further models \cite{FaceFormer, Imitator} build on this line of work using transformers to great success. However, while these models capture variation in identity, they do not model motion not correlated to audio, such as that caused by emotion. Several other regression-based methods \cite{thies2020nvp, LiveSpeechPortraits, wen2020photorealistic, LipSync3D, Tang2022} attempt to predict animation from audio. However, as these are regression-based, the motion that is not explicitly modeled is averaged out in the predictions. This leads to muted upper head motion, no eye motion, and, unless specifically modeled, no emotion. Our method, being GAN-based and using ground truth data as a driving signal, does not suffer from this problem.

\section{Method}

\begin{figure*}
    \centering
    \includegraphics[width=\textwidth]{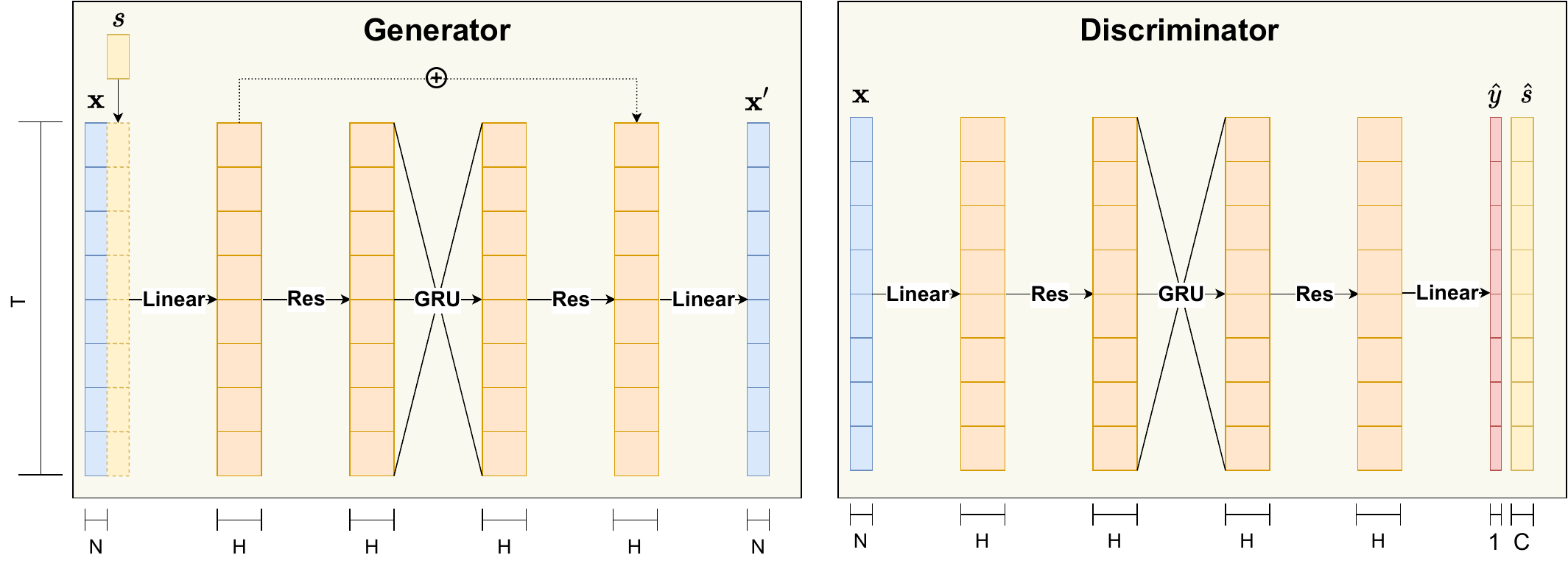}
    \caption{A diagram showing the architecture for both the generator and discriminator networks. Here $\mathbf{x}$ and $\mathbf{x'}$ are sequences of Metahuman rig controls, $s$ is a style label, and $y$ is a prediction of if the sequence is real or fake. Each network consists of linear layers, residual feed-forward layers, and a bi-directional GRU.}
    \label{fig:Architecture}
\end{figure*}

Given an animation $\mathbf{x}$, represented as a sequence of MetaHuman rig controls, our proposed method aims to generate a new animation $\mathbf{x'}$ with the desired style $s'$ while preserving the content of $\mathbf{x}$. We adopt the StarGAN framework \cite{StarGAN2018} as the basis of our method, which consists of a generator network $\mathcal{G}$ and a discriminator network $\mathcal{D}$. The generator network takes both the input animation $\mathbf{x}$ and the target style label $s'$ as input, and outputs the new animation $\mathbf{x'} = \mathcal{G}(\mathbf{x}, s')$. The discriminator network has two parts: a critic part $\mathcal{D}_{crt}$ that distinguishes between real and fake data, and a classification part $\mathcal{D}_{cls}$ that predicts the style label of the input data. Specifically, the critic part outputs a scalar value $\hat y = \mathcal{D}_{crt}(\mathbf{x})$ that measures the realism of the input data, while the classification part outputs a probability distribution $\hat s = \mathcal{D}_{cls}(\mathbf{x})$ over all possible style labels.

To ensure that our generated animations are temporally consistent, we need to consider the temporal dynamics of the animation data. Therefore, we represent our data as a time series by concatenating the per-frame vectors of animation controls over a temporal window of length $T$. This results in sequential data of the form $x \in \mathbb{R}^{T \times N}$, where $N$ is the dimensionality of the per-frame vector.

Our proposed network is a modification of the StarGAN, altered to work for sequential animation data. We first discuss the changes we make to the architecture to make this possible. Then we introduce the losses required for animation style transfer, including a novel viseme matching loss.  

\subsection{Network Architecture}

\begin{wrapfigure}{l}{0.5\columnwidth}
    
    \includegraphics[width=0.45\columnwidth]{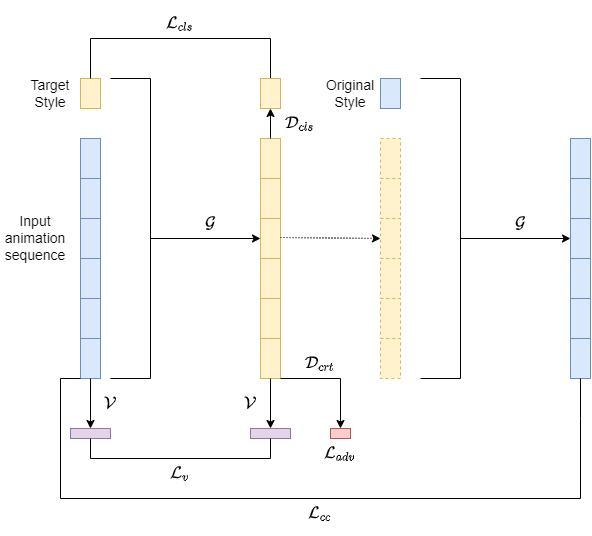}
    \caption{Overview of the multiple losses used for training this model. Here $\mathcal{G}$ and $\mathcal{D}$ are the generator and discriminator networks to be trained and $\mathcal{V}$ is the pre-trained viseme classifier.}
    \label{fig:Losses}
\end{wrapfigure}

The generator and discriminator share similar high-level architectures. Each consists of an encoder that maps per-frame rig controls to a higher-dimensional latent space, a recurrent layer to incorporate temporal information, and a decoder to map from the latent space to the desired domain. The encoders and decoders both make use of residual layers \cite{ResNet16} and dropout \cite{Dropout14}.

The generator takes as input a sequence of rig controls $\mathbf{x} \in \mathbf{T \times N}$ and a categorical style code $s' \in \{ 0, 1\}^C $, where any present styles are represented as a $1$, and outputs a new sequence. This is done by first repeating the style code over the time axis and concatenating it to the sequence (see Figure {\ref{fig:Architecture}}). This concatenated sequence is then fed through the encoder and recurrent layers. The generator contains a skip connection between the first and last hidden layers to encourage the conservation of content and improve convergence. The final layer is a linear layer to project the latent representation back into the rig parameter space.

The discriminator is similar but it does not take a style code and has no skip connection. The final linear layer produces both per-frame probabilities that the input is real and per-frame style codes. Both networks use a multi-layer, bi-directional GRU \cite{GRU14} as the recurrent layer. These architectures are shown in Figure \ref{fig:Architecture}.

\subsection{Losses}

The networks each have a separate combination of loss functions. These are given as follows:

\begin{equation}
    \mathcal{L}_{\mathcal{G}} = \lambda_{cc}\mathcal{L}_{cc} +  \lambda_{cls}\mathcal{L}^G_{cls} +  \lambda_{v}\mathcal{L}_{v} - \lambda_{adv}\mathcal{L}_{adv}
\end{equation}
\begin{equation}    
\mathcal{L}_{\mathcal{D}} = \lambda_{cls}\mathcal{L}^D_{cls} + \lambda_{adv}\mathcal{L}_{adv}
\end{equation}

\subsubsection{Cycle Consistency}

With facial animations, it is nearly impossible to get paired data, as this would require a frame-to-frame correspondence between animations of different styles. The cycle consistency loss, first introduced for the CycleGAN \cite{CycleGAN2017} enables us to overcome this barrier by using unpaired data. We can apply a new style $s'$ to an animation $\mathbf{x}$ to obtain $\mathbf{x'}$ and then reapply the original style code $s$ to form a cycle. The cycle consistency loss is then the difference between $\mathbf{x}$ and the cycle. This loss encourages the generator to only change attributes associated with style, and to preserve the content as much as possible.

\begin{equation}
    \mathcal{L}_{cc} = \left|\left| \mathbf{x} - \mathcal{G}(\mathcal{G}(\mathbf{x}, s'), s)\right|\right|_2
\end{equation}

Here $\left|\left| \cdot \right|\right|_2$ denotes the $L_2$ norm, the use of which is a modification from the image-based starGAN, which uses $L_1$ as we find animation data is noisy, and it is important to hit extremes of motion for lip-sync.

\subsubsection{Classification Loss}

Given an animation sequence $\mathbf{x}$ and style label $s'$ we want the generator to produce a new animation $\mathbf{x'}$ that has the style characteristics of $s'$. We are able to do this using the classification branch of $\mathcal{D}$. This classifier should be capable of correctly identifying the style code $s'$. Simultaneously, the generator should be encouraged to produce sequences that the classifier identifies as having the desired style. This means we must decompose the classification loss $\mathcal{L}_{cls}$ into two components. The first of these components is used to train $\mathcal{D}_{cls}$ to correctly label the styles of the real training data. This is given by the cross entropy loss:

\begin{equation}
    \mathcal{L}_{cls}^{D} = -\mathbb{E}_{\mathbf{x}, s}\left[ s \log \mathcal{D}_{cls}(\mathbf{x}) \right]
\end{equation}

This loss is calculated for real sequences only. The second component of this loss is for training $\mathcal{G}$ to produce sequences that are labeled correctly. This loss is of the form:

\begin{equation}
    \mathcal{L}_{cls}^{G} = -\mathbb{E}_{\mathbf{x'}, s'}\left[ s' \log \mathcal{D}_{cls}( \mathcal{G}(\mathbf{x}, s')) \right]
\end{equation}

\subsubsection{Adversarial Loss}

In addition to creating animations with a desired style, the generator needs to produce animations that appear realistic. This is achieved through an adversarial loss. We do not use the adversarial loss first defined for GANs \cite{GAN14}, but a Wasserstein loss with gradient penalty \cite{WGAN17, WGANGP17}. Such a loss has been shown to improve the stability of the GAN during training. Here $\mathcal{L}_{gp}$ is the gradient penalty and $\lambda_{gp}$ is a weight controlling the influence its influence. We use $\lambda_{gp} = 10$ for our experiments.

\subsubsection{Preserving Lip-Sync}

Any animation style transfer method must preserve the motion of the lips in such a way that they remain synchronised to audio. This is a challenging task as even small errors can lead to the lips appearing out of sync. Previous work has \cite{NSPVD} addressed this by using a cosine mouth loss to preserve the general shape of the mouth. However, such a method will not work in our case as we require the shape of the mouth to vary with style. For example, applying a happy style should be expected to pull the corners of the mouth into a smile. 

Ideally, we would use the audio to encourage lip sync. However, we do not alter the style of the audio, so any naive attempt to match animation to audio would work against the style transfer. Specifically, consider an expert discriminator trained with contrastive learning, as is common in the 2D domain \cite{Wav2Lip, LipGAN, Liang22, Song22, wang2023seeing, shen2023difftalk, stypulkowski2022diffused}. If we were to train such a model using stylised data, all in-sync pairs would have the same style. Such a discriminator would then only recognise data as in-sync if it had the same style. Say, for example, we then wanted to convert happy to sad and used this expert discriminator, the expert loss would encourage the model to output happy animation to maintain it's notion of sync. To overcome this limitation, we need a style-independent representation of speech. For this we use visemes \cite{PhonemeVisemeMap}, the visual counterpart to phonemes. The goal produce a network that predicts visemes from animation which can be used as an additional loss for the generator. 

To do this, we first use a pre-trained phoneme classifier \cite{Phoneme21, HuggingFace} to predict phonemes from audio. This model is based on Wav2Vec2 \cite{wav2vec220} and finetuned on the dataset common voice \cite{commonvoice20} to predict phonemes. This classifier takes as input raw waveforms and outputs unnormalised log probability for $392$ phoneme tokens over temporal windows of $0.02s$. We then resample the outputs to 60Hz using linear interpolation. This gives us phoneme probabilities for each frame in our animations. Multiple phonemes are expressed with the same mouth shape, so we use a phoneme-to-viseme map \cite{PhonemeVisemeMap} to obtain per-frame predictions of $16$ visemes from audio. 

Next we train a neural network classifier $\mathcal{V}: \mathbb{R}^{T \times N} \rightarrow \mathbb{R}^{T \times 16}$ to predict the visemes from the animation curves alone. Such a network has a few advantages. The classifier takes in only animation data and does not require audio, therefore it can be applied to animations created by the generator which do not have corresponding audio. It also predicts visemes \textbf{independent} of speaker identity and emotion, giving a fixed measure of lip shape that can be preserved by the generator. The network has a similar architecture to the discriminator network, with an encoder, GRU layer, and decoder. The only difference is that the final linear layer has different output dimensions to match the number of visemes.

We can then use this network to derive the \textit{viseme-preserving loss}, which is simply the cross-entropy between the viseme classifications of the real sequences $\mathbf{x}$ and the viseme classifications of the generated sequences $\mathcal{G}(\mathbf{x}, s')$. The loss encourages the generator to produce sequences with the same mouth shapes as the input sequence, without constraining the style.

\section{Results}

\subsection{Data}

\begin{wrapfigure}{l}{0.5\columnwidth}
    \centering
    \includegraphics[width=0.45 \columnwidth]{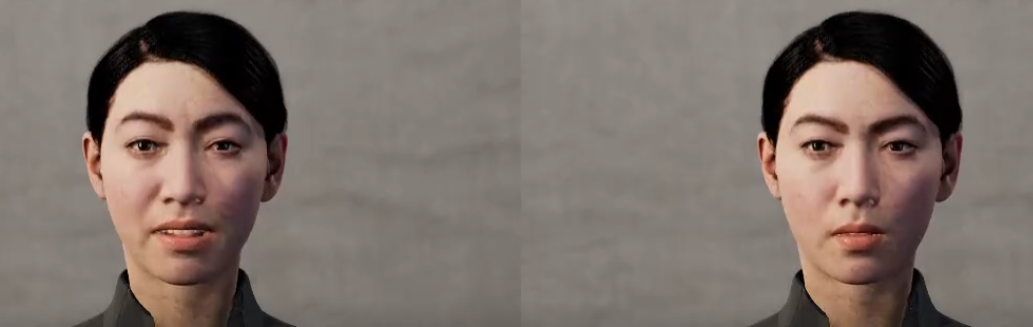}
    \caption{While less easy to quantitatively measure than emotion, our method also handles idiosyncratic style. Here we show the same animation in the style of two different actors.}
    \label{fig:Idiosyncratic}
\end{wrapfigure}

Our dataset consists of a total of 30 minutes of animations across two custom MetaHuman-based rigs \cite{MetaHumans}. The animations are obtained using a head-mounted camera worn by two professional actors. The video footage is then tracked and solved to custom rigs at 60 fps using proprietary software. A subset of $71$ controls is used for our dataset. Each actor is asked to record a series of performances consisting of spoken sentences in the desired emotion. Three emotions are chosen: happy, neutral, and sad. For each of these emotions, we record approximately 5 minutes of animation per actor. In order to ensure we have captured a wide enough range of facial motion, we make use of seven phonetic pangrams. These sentences contain many different phonemes. Each is recorded twice, along with rest periods where the emotion is held without any speech. This gives us a total of 6 different forms of style, which is sufficient to demonstrate that our method can indeed perform many-to-many style transfer.

We note that while the closest similar method to ours \cite{NSPVD} is tested on 3DMM parameters obtained using monocular reconstruction on real video, we have chosen instead to use professional performance capture. Such data is much more realistic and allows us to better capture the nuance of emotional and idiosyncratic styles.

\subsection{Implementation Details}

We implement the proposed method using pytorch \cite{Pytorch}. For all networks, we use a hidden dimension of $256$. Dropout is applied after each layer, except the GRU, with probability $0.4$. The networks are trained with a batch size of $32$ and using the Adam optimiser \cite{Adam15} with learning rate of $10^{-4}$. During training, the sequence length is fixed to 30 frames. During inference, the sequence length can be arbitrary. We train for $100$ epochs taking approximately 12 hours on an NVIDIA Quadro P6000. We also apply data augmentation such as time stretching and the addition of small amounts of noise to the audio features.

\subsection{Metrics}

In order to justify the improvement made by our work, we quantitatively and qualitatively evaluate the results. For lip sync, a major focus of our work, quantitative metrics do exist. We use a pre-trained syncnet \cite{Chung16a} to measure lip sync by passing Maya \cite{Maya} renderings of the animations, together with the corresponding audio. We report the LSE-C and LSE-D metrics as proposed in Wav2Lip \cite{Wav2Lip}. The LSE-D metric quantifies the level of lip-sync, while the LSE-C metric measures the confidence, where low scores for LSE-C mean that the audio and video are only weakly correlated.

\begin{figure}
    \centering
    \includegraphics[width=0.8\columnwidth]{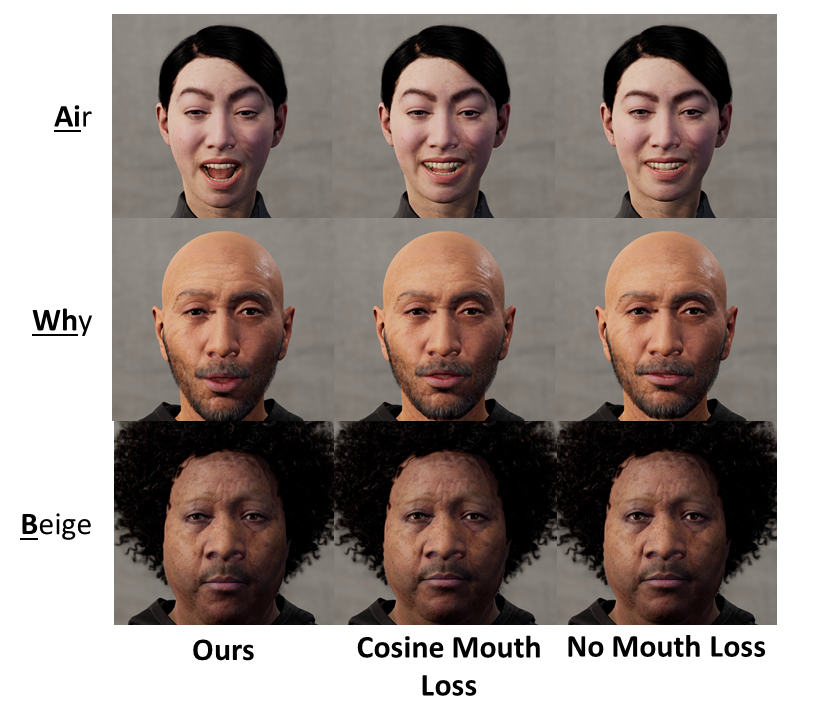}
    \caption{The presence of the Viseme Loss in our work significantly improves the lip sync when compared to both our method without this loss, and the work of \cite{NSPVD}}
    \label{fig:LipSync}
\end{figure}

\begin{wraptable}{r}{0.6\columnwidth}
    \centering
    \begin{tabular}{c|c c c} 
    
     \hline
     Method & LSE-D $\downarrow$ & LSE-C $\uparrow$ &  Emo-P $\uparrow$ \\ 
     \hline
     Ours w/o Viseme Loss & 9.501 & 1.781 & 0.421 \\
     Cosine Mouth Loss \cite{NSPVD} & 8.699 & 2.667 & 0.417 \\
     \textbf{Ours Full} & \textbf{7.33} & \textbf{4.667} & \textbf{0.443} \\
     \hline
     Captured Data & 7.617 & 4.502 & 0.385 \\
      \hline
    \end{tabular}
    \caption{A table comparing our work with and without the Viseme Loss to that of Neural Style Preserving Visual Dubbing \cite{NSPVD} across metrics for lip sync (LSE-C and LSE-D) and for emotional clarity (EMO-P).}
    \label{tab:main_res_table}
\end{wraptable}

In our experiments, we perform two forms of style transfer. Idiosyncratic style transfer is difficult to quantify. However, figure \ref{fig:Idiosyncratic} demonstrates that our model does indeed learn to alter this. For emotional style transfer, we can define a quantitative metric. For this purpose, we use a pre-trained emoNet \cite{emonet2021estimation}. This network predicts the per-frame emotion of a given video without considering the audio, outputting logits related to each of the five major emotions. We restrict the output of this network to just the three emotions we are considering, that is happy neutral, and sad, and take the softmax to give pseudo-probabilities for each emotion. We propose using this as a metric, where the value of this metric is the pseudo-probability associated with the desired emotion after style transfer. We denote this metric as Emo-P.

\subsection{Comparisons}

\textbf{Quantitative:} To the best of our knowledge, the only work to attempt style transfer for facial animation is the work of Neural Style-Preserving Visual Dubbing (NSPVD) \cite{NSPVD}. Our work varies from theirs in two key ways. The first is that we use a starGAN whereas they use a CycleGAN. These networks perform different tasks and it is therefore difficult to compare the two. The second is our use of the viseme preserving loss in place of a cosine mouth loss. We compare the effects of this novel loss in table \ref{tab:main_res_table}. In addition to comparing our work to NSPVD \cite{NSPVD}, we also perform a simple ablation study to demonstrate that the inclusion of the viseme preserving loss does improve the lip-sync. These results are also in table \ref{tab:main_res_table}. Our method outperforms the mouth cosine loss from NSPVD \cite{NSPVD} across all metrics, which in turn is an improvement over not using any additional mouth loss. Interestingly, our proposed viseme loss improves the lip sync even compared to the ground truth. This may be because the model exaggerates key lip shapes. It is of interest to future work to determine if this loss could improve performance capture animation where the lip sync is poor.

\begin{wraptable}{r}{0.6\columnwidth}
    \centering
    \begin{tabular}{c|ccc}
        \hline
         Statement & Lip-sync & Naturalness & Emotion \\
         \hline
         Ours $>$ Cosine Mouth Loss & 63 & 76 & 83 \\
         Ours $>$ No Mouth Loss & 67 & 62 & 68 \\
         \hline
    \end{tabular}
    \caption{The results of our user study (N = 10), showing the percentage of users that agree with the provided statement.}
    \label{tab:user_study}
\end{wraptable}

\textbf{Qualitative:} We are also able to show the effectiveness of our method qualitatively. Figure \ref{fig:LipSync} shows example frames of animation. It can be seen that our method produces the expected lip shapes for the highlighted section of the given words better than NSPVD \cite{NSPVD} and the baseline method. For example, the start of the word ``why" has more clearly pursed lips. For further qualitative results, we refer to the supplementary material.

\textbf{User Study:} We also run a two-alternative forced choice test to compare our method with the cosine mouth loss \cite{NSPVD} and with no mouth loss. The results are shown in table \ref{tab:user_study} and show that our method is preferred across all metrics. Of particular note is how emotion is much clearer in our method compared with the cosine mouth loss, validating our hypothesis that the mouth is over-constrained.

\section{Conclusions and limitations}

In this work, we proposed a novel method for style transfer in facial animation using the StarGAN framework and a viseme preserving loss. Our method is able to generate new facial animations with a desired style label while preserving the content of the original animation. We also demonstrated that the inclusion of the viseme preserving loss improves lip-sync in the generated animations. Our method is able to perform style transfer across multiple styles and produce temporally consistent animations. Compared to previous work on facial animation style transfer, our method achieves superior results in terms of lip-sync accuracy and style transfer quality. We believe that our method has potential applications in areas such as animation and film production, where the ability to quickly and easily modify the style of facial animations while preserving content can be highly beneficial. Future work could focus on improving the generalization of the method to different datasets and exploring the use of additional losses to further improve the quality of the generated animations.

\bibliographystyle{plain}
\bibliography{bibliography}

\end{document}